\documentstyle[aps,prd,twocolumn,axodraw,epsfig]{revtex}

\def\stacksymbols #1#2#3#4{\def\theguybelow{#2}
    \def\vp{\lower#3pt}
    \def\sp{\baselineskip0pt\lineskip#4pt}
    \mathrel{\mathpalette\intermediary#1}}

\def\intermediary#1#2{\vp\vbox{\sp
     \everycr={}\tabskip0pt
     \halign{$\mathsurround0pt#1\hfil##\hfil$\crcr#2\crcr
              \theguybelow\crcr}}}
\def\beq{\begin{equation}}
\def\eeq#1{\label{#1}\end{equation}}
\def\eeqn{\end{equation}}
\def\beqa{\begin{eqnarray}}
\def\eeqa#1{\label{#1}\end{eqnarray}}
\def\eeqan{\end{eqnarray}}

\def\leqn#1{(\ref{#1})}

\def\({\left(}
\def\){\right)}
\def\to{\rightarrow}

\def\gapproxeq{\stacksymbols{>}{\sim}{2.5}{.2}}

\newcommand{\bspace}{\!\!\!\!}
\def\met{\mbox{$E{\bspace}/_{T}$}}

\begin{document}

\wideabs{

\title{Collider Signature of T-quarks}
\author{Marcela Carena$^{1}$, Jay Hubisz$^{1}$, Maxim Perelstein$^2$, and
Patrice Verdier$^{3}$}
\address{
$^1$Fermi National Accelerator Laboratory, Batavia, Illinois 60510, USA\\
$^2$ Laboratory of Elementary Particle Physics, Department of Physics,
Cornell University, Ithaca, New~York~14853, USA\\
$^3$ Institut de Physique Nucl\'eaire de Lyon, IN2P3-CNRS,
Universit\'e Lyon 1, Villeurbanne, France
}

\date{\today}
\maketitle

\begin{abstract}
Little Higgs models with T~Parity contain new vector-like fermions, the  
\mbox{T-odd} quarks or ``\mbox{T-quarks}'', which can be produced at hadron colliders
with a QCD-strength cross section. Events with two acoplanar jets and large 
missing transverse energy provide a simple signature of \mbox{T-quark} production. 
We show that searches for this signature with the Tevatron Run II data
can probe a significant part of the Little Higgs model parameter space not
accessible to previous experiments,
exploring \mbox{T-quark} masses up to about 400~GeV.  This reach covers
parts of the
parameter space where the lightest \mbox{T-odd} particle can account for the
observed dark matter relic abundance.  We also comment on the
prospects for this search at the Large Hadron Collider (LHC).
\end{abstract}
}

{\em Introduction ---} Little Higgs (LH) models~\cite{littlest} (for reviews, 
see~\cite{LHrev1,LHrev2}) provide an interesting scenario for physics at the
TeV scale, alternative to other popular scenarios such as supersymmetry or
extra dimensions. The LH models contain a Higgs 
boson of mass $m_h$ up to several hundred GeV, 
as well as additional gauge bosons, fermions,  
and scalar particles with masses in the 100~GeV -- 5~TeV range. These models 
describe the physics up to a ``cutoff scale'' of order 10 TeV, beyond which
they need to be embedded in a more fundamental theory. The hierarchy 
between the Higgs mass and the cutoff scale is due to the fact that the
Higgs is a pseudo-Nambu-Goldstone boson, corresponding to a global symmetry 
spontaneously broken at a scale $f\sim 1$ TeV. Explicit breaking
of the global symmetry by gauge and Yukawa couplings induces a non-trivial 
Higgs potential via quantum effects, triggering electroweak symmetry breaking
(EWSB). However, the one-loop quadratically 
divergent contribution to the Higgs mass vanishes due to the special
``collective'' nature of the explicit global symmetry breaking, and thus the 
Higgs mass can be achieved with minimal fine-tuning.

Early implementations of the Little Higgs mechanism suffered from severe 
constraints
from precision electroweak fits~\cite{PEW}. These constraints are elegantly 
avoided by the introduction of T~Parity~\cite{LHT}, a discrete ${\cal Z}_2$ 
symmetry which is constructed in such a way that all the Standard Model (SM) 
states are even, while most new \mbox{TeV-scale} states of the LH model are odd.
This construction forbids all tree-level corrections to precision electroweak 
observables from the new states. Many LH models can be extended to 
incorporate T~Parity. In this letter, we focus on one of the simplest 
examples, the 
Littlest Higgs model with T~Parity (LHT)~\cite{Low}. Precision electroweak
constraints on this model have been analyzed at the one-loop level~\cite{HMNP},
and it was shown to provide consistent fits for values of $f$ as low as 
500~GeV, 
avoiding fine tuning. The model also provides an attractive dark matter 
candidate~\cite{HM,BNPS}, the lightest \mbox{T-odd} particle (LTP).  

Consistent implementation of T~Parity in the LHT model requires the 
introduction of a \mbox{T-odd} vector-like fermion partner for each
left-handed T-even SM 
fermion, yielding
six new Dirac \mbox{T-odd} quarks (``\mbox{T-quarks}''),
$\tilde{Q}_i^a=(\tilde{U}_i^a, \tilde{D}_i^a)$, and six new Dirac
\mbox{T-odd} leptons (``\mbox{T-leptons}''),
$\tilde{L}_i=(\tilde{E}_i,\tilde{N}_i)$,
where $i=1\ldots 3$ is
the generation index and $a$ is the color index.  The masses 
of these states lie in the 100~GeV -- few~TeV range. The main goal of 
this letter is to analyze the collider phenomenology of the \mbox{T-quarks}. 
 
{\em Model ---} The LHT model has been discussed in detail 
elsewhere~\cite{LHrev2,Low,HMNP,HM}; here, it suffices to summarize the 
features important for our analysis. The LHT is based on a weakly gauged 
non-linear 
sigma model (nlsm). The global symmetry breaking pattern is $SU(5)\to SO(5)$, 
resulting in 14 Nambu-Goldstone bosons. The symmetry breaking scale, $f$
(the pion decay constant of the nlsm), is a free parameter; in the absence of 
fine tuning, $f\sim 1$ TeV. The nlsm is valid up 
to the cutoff scale $\Lambda\sim 4\pi f \sim 10$ TeV. The physics above the 
cutoff scale will not be discussed
here since it is outside of the reach of the Tevatron and the LHC.
The gauge symmetry group is $[SU(2)\times U(1)]^2$, broken at the
scale $f$ down to the diagonal subgroup, $SU(2)_L\times U(1)_Y$, which is 
identified with the SM electroweak gauge symmetry. Four Nambu-Goldstone 
bosons are absorbed in this breaking; the remaining 10 form the SM Higgs doublet, $H$, and a new $SU(2)_L$ scalar triplet, $\Phi$.

At the
quantum level, explicit breaking of the $SU(5)$ global symmetry by gauge and 
Yukawa couplings induces a potential for $H$ and triggers EWSB,
$SU(2)_L\times U(1)_Y\to U(1)_{\rm em}$. The action of T~Parity in the  
gauge sector interchanges the two sets of $SU(2)\times U(1)$ gauge fields. The T-even
combinations of the gauge bosons correspond to the SM $W, Z$ and $\gamma$,
whereas the \mbox{T-odd} combinations, denoted 
by $\tilde{W}^a (a=\pm,3)$ and $\tilde{B}$, acquire masses at the scale $f$:
\beq
M(\tilde{W}^a)\,\approx\, gf,~~~M(\tilde{B})\,\approx\, 
\frac{g^\prime f}{\sqrt{5}}\approx 0.16 f,
\eeq{mass}
where $g$ and $g^\prime$ are the SM $SU(2)_L$ and $U(1)_Y$ gauge couplings,
and corrections of order $v^2/f^2$ due to the EWSB have been neglected.
Note that the \mbox{T-odd} $U(1)$ gauge boson, $\tilde{B}$, or so called
``heavy photon'' 
is quite light compared 
to $f$. In large parts of the parameter space, the $\tilde{B}$ is the
LTP, and is stable. Since the $\tilde{B}$ is weakly
interacting, its  
stability poses no cosmological difficulties, and in fact it can act as WIMP 
dark matter~\cite{HM,BNPS}. In the analysis of this letter we will assume that
the heavy photon is the LTP. The scale $f$ is bounded from below by precision
electroweak data~\cite{HMNP} and the corresponding bound on the LTP mass is 
$M(\tilde{B})>80$~GeV.

The masses of the \mbox{T-quarks} and 
\mbox{T-leptons}
are given by
\beq
M_{ij}(\tilde{Q}) \,=\, \kappa_{ij}^Q f,~~~
M_{ij}(\tilde{L}) \,=\, \kappa_{ij}^L f,
\eeq{tmasses}
where the couplings $\kappa$ are free parameters. 
In this letter, we will focus on the \mbox{T-quarks} of the first two generations, 
and assume that they have a common mass,
$\tilde{M}$. This degeneracy eliminates any potential loop-level 
flavor-changing effects via the GIM mechanism~\cite{LHTflavor}. Experimental bounds on the 
flavor-conserving four-fermion operators such as $eeuu$ and $eedd$ imply 
the bound~\cite{HMNP}
\beq
\tilde{M} < 4.8~{\rm TeV} \left(\frac{f}{\rm TeV}\right)^2\,.
\eeq{upperT}
T-quark contributions to precision electroweak observables
have been computed in~\cite{HMNP}, and do not impose any new bound on $\tilde{M}$.
To avoid charged/colored LTP, we require \mbox{$\tilde{M}>M(\tilde{B})$}.

The LHT model contains additional states in the top sector, required to
cancel the one-loop quadratic divergence in the Higgs mass from top loops.
The collider phenomenology of these states 
~\cite{HM,top}, however, does not play a role in this analysis.

Before proceeding, it is useful to compare and contrast the spectrum of 
the LHT model with the
more familiar case of the minimal supersymmetric standard model (MSSM). In
both models, SM states acquire parity-odd partners with the same gauge
quantum numbers. For example, the $\tilde{W}^a$ and $\tilde{B}$
  bosons of the LHT model  
are the 
analogues of the wino and bino of the MSSM; the \mbox{T-quarks} and \mbox{T-leptons} are the 
counterparts of squarks and sleptons. The two important differences are:
(1) the LHT partners have the same spin as the SM states; and (2) in the 
LHT, partners only exist for a subset of the SM: for example, the 
right-handed SM fermions and the gluon do not acquire \mbox{T-odd} partners.

{\it Collider Signatures ---} At a hadron collider, the \mbox{T-quarks} can be 
pair-produced via QCD processes:
\beq
q\bar{q}\to\tilde{Q}_i\bar{\tilde{Q}}_i,~~~gg\to\tilde{Q}_i\bar{\tilde{Q}}_i.
\eeq{process} 
The produced \mbox{T-quarks} decay promptly. Due to conserved T~Parity, their decay
products necessarily contain the LTP $\tilde{B}$, leading to a missing energy
signature in the detector. In particular, the decay channel
\beq
\tilde{Q}_i \to q_i \tilde{B}
\eeq{decay}
%
\begin{figure}[t!]
\begin{center}
\includegraphics[width=8.5cm]{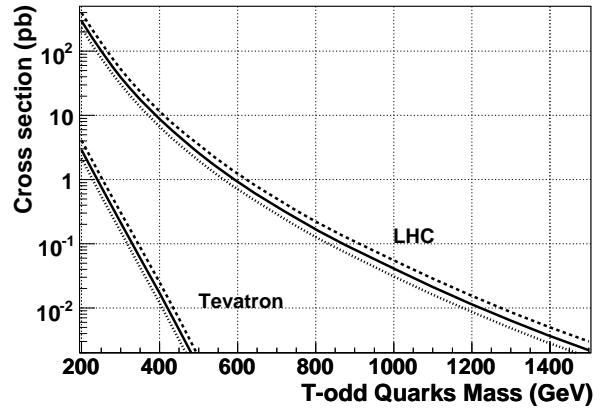}
\vskip4mm
\caption{Cross section of \mbox{T-quark} pair production (per flavor) at 
the Tevatron Run~II and at the LHC. Solid, dashed and dotted lines
correspond to $\mu=\tilde{M}, \tilde{M}/2$ and $2\tilde{M}$,
respectively.}
\label{fig:xsec}
\end{center}
\end{figure}
\noindent is open throughout the parameter space for $q_i=u,d,s, 
c$, with the exception of very  narrow
bands where the \mbox{T-quarks} and the LTP are nearly-degenerate. Events with
both \mbox{T-quarks} decaying in this channel result in a $2j+\met$ signature
with acoplanar jets both
at the Tevatron 
and the LHC. Within the \mbox{T-quark} mass range accessible at the Tevatron, 
the branching ratio in the channel~\leqn{decay} is very nearly 100\%.
For heavier \mbox{T-quarks}, competing channels such as $q \tilde{W}$ may 
open up, which could be relevant for LHC studies.

To analyze the experimental reach in terms of the model parameters, we have 
implemented the relevant sector of the LHT model in the 
{\tt MadGraph}~\cite{madgraph} parton-level event generator and simulated 
the reaction~\leqn{process},~\leqn{decay}. The total \mbox{T-quark} production cross 
sections (per \mbox{T-quark} flavor) at the Tevatron Run~II and the LHC are shown in 
Fig.~\ref{fig:xsec}. The CTEQ6L1 PDF set~\cite{cteq} was used, and 
renormalization and factorization scales, $\mu$, 
were varied between $\tilde{M}/2$ and $2\tilde{M}$ to obtain an estimate of
the associated uncertainty. The rather large uncertainty (typically about 
30\%) is primarily due to the use of
the leading-order matrix element, and could be improved by a 
next-to-leading order calculation of the process~\leqn{process} in the
LHT model. 
Based on the studies of squark production processes with similar 
kinematics, we expect that the NLO cross section is enhanced by $K\sim 1.3$
compared to the LO estimate.  However, we do not rescale our leading
order result, and so we expect our estimate is conservative.


The counterpart of the process~\leqn{process},~\leqn{decay} in the MSSM is the production of squark pairs followed by the decay
$\tilde{q}\to q\tilde{\chi}_1^0$. The production cross section of \mbox{T-quark} 
pairs 
is larger than that of squarks with the same mass due to the spin sum
of the final state. However, if the \mbox{T-quark} and squark masses, as well
as the LTP and LSP masses
are equal, we find that the properties of the final-state jets
(e.g. transverse energy and rapidity distributions) are essentially identical. 
Therefore, with the appropriate overall rescaling, pair production 
and decay of \mbox{T-quarks} can be perfectly simulated using
PYTHIA~6.323~\cite{Pythia} as an MSSM event generator which goes beyond
the parton level.
For this analysis, we generate a set of Monte Carlo events equivalent to the pair production of 
first and 
second generation \mbox{T-quarks}, assuming 100\% branching fraction for the
decay channel~(\ref{decay}).


The D\O\ experiment developed two analyses searching for events in the 
acoplanar dijet topology using 310~$pb^{-1}$ of data recorded during the 
Tevatron Run~II, which can be reinterpreted as \mbox{T-quark} searches.
The first one (``analysis A'') is the search for squark pair
production described in~\cite{D0squark}. This analysis is efficient
for large mass 
differences between the \mbox{T-quarks} and the LTP. The second analysis
(``analysis B'')
is the search 
for scalar leptoquarks decaying into a quark and a neutrino~\cite{D0lq}, 
which is more efficient for low mass differences between \mbox{T-quarks} and the LTP. 
Those two event selections were applied to the Monte Carlo samples described 
above to extract signal efficiencies. 
As those Monte Carlo samples do not have any detector simulation, the signal 
efficiencies obtained in this way are overestimated. To take that effect into 
account, the signal efficiencies obtained at the generator level were compared
to those reported in~\cite{D0squark,D0lq}. A conservative scale factor of 0.75 
was applied on all \mbox{T-quark} signal efficiencies.

\begin{figure}[t!]
\begin{center}
\includegraphics[width=8.5cm]{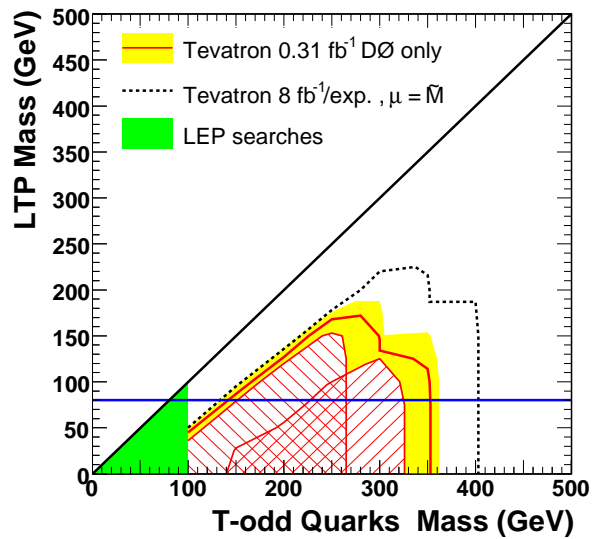}
\vskip2mm
\caption{Present and projected reach of the Tevatron Run~II search for
  \mbox{T-quarks}  
in the ($\tilde{M}$,$M(\tilde{B})$) plane. The thick red line shows the expected
excluded region at the 95\% C.L. for a luminosity of 310~$pb^{-1}$, and $\mu = \tilde{M}$.
The yellow band 
shows the 
effect of varying the renormalization and factorization scale between 
$\mu=2\tilde{M}$ and $\mu=\tilde{M}/2$.  
The hatched regions show the expected excluded regions by analysis~A and B 
separately, for $\mu=2\tilde{M}$.
The dotted black line shows the projection to an integrated 
luminosity of 8~$fb^{-1}$ per experiment at the Tevatron for $\mu=\tilde{M}$. 
The thick blue line corresponds to the indirect lower limit on the $\tilde{B}$ 
mass from precision electroweak data. 
}
\label{fig:exclu}
\end{center}
\end{figure}

In Figure~\ref{fig:exclu}, we present the expected mass limits at the
95\%~C.L. in the ($\tilde{M}$,$M(\tilde{B})$) plane.  We have used
the number of expected background events 
and the systematic uncertainties reported in~\cite{D0squark,D0lq}, as
well as the leading order \mbox{T-quark} pair production cross sections from
Figure~\ref{fig:xsec}.  In addition, the limits are computed using the
modified frequentist $CL_s$ method~\cite{CLS}.
Also shown in Figure~\ref{fig:exclu} are the regions excluded by the
precision electroweak data, which 
place a lower bound on the scale $f$ of about 500~GeV (see Ref.~\cite{HMNP}
for details) corresponding to $M(\tilde{B})\gapproxeq 80$~GeV, and by the LEP
squark searches~\cite{LEPsquark}. Note that the LEP reach for squarks is
limited by kinematics, so that the reach for \mbox{T-quarks} is
nearly identical 
in spite of different production cross sections.  With only 310~$pb^{-1}$, 
Tevatron Run~II data can place relevant 
bounds on the \mbox{T-quark} and LTP masses, probing a region of the parameter 
space not accessible to previous experiments. Taking into account the 
factorization scale
uncertainty, the expected lower bound on the \mbox{T-quark} mass is 
approximately 325~GeV 
if $\tilde{M}-M(\tilde{B})\gapproxeq 245$~GeV (where analysis A is applicable)
and 265~GeV if  $\tilde{M}-M(\tilde{B})\gapproxeq 140$~GeV (where 
analysis B is used). There is no strict bound
for smaller values of the T-quark-LTP mass difference, since in this case the 
produced jets are too soft to be detected. The reach 
can be extended further with additional integrated luminosity. An 
extrapolation to 8~$fb^{-1}$/experiment shows that \mbox{T-quark} masses up to 400~GeV 
will be probed (see Fig.~\ref{fig:exclu}). 
 
A search for the $2j+\met$ signature at the LHC is expected to have 
significantly better reach in $\tilde{M}$ due to the higher \mbox{T-quark} 
production cross sections, see Fig.~\ref{fig:xsec}. We estimate that 
the \mbox{T-quark} masses up to about 850~GeV could be probed at the 
3$\sigma$ level with a few fb$^{-1}$ of integrated luminosity. To obtain
this estimate, we computed the number of signal and background events at the 
parton level, imposing the cuts $\met\geq 200$~GeV, $p_{T~{\rm jet}}\geq 
200$~GeV. We assumed that with these cuts the background is dominated by
the irreducible component, $Zjj$ with the $Z$ decaying 
invisibly. This background can be calibrated using the events with the $Z$ 
decaying leptonically; we assumed that the accuracy of this calibration is 
10\%. Note, however, that instrumental backgrounds, such as pure
QCD multi-jet events with apparent $\met$ due to jet mismeasurement, will
likely play an important role in limiting the reach. A careful analysis of 
this issue, including a full detector simulation, is required to 
obtain a more robust estimate of the reach. Note also that our analysis 
assumed Br$(\tilde{Q}\to q\tilde{B})=100$\%, and is not directly applicable
if $M(\tilde{W})<\tilde{M}$.
(For a recent discussion of the 
potential signatures of \mbox{T-quarks} in leptonic final states at the
LHC, see~\cite{Freitas,Belyaev}.)

%
%



\begin{figure}[t!]
\begin{center}
\includegraphics[width=8.5cm]{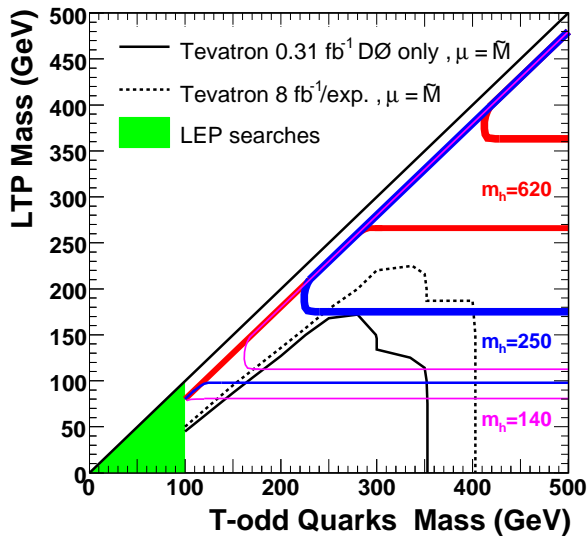}
\vskip2mm
\caption{Prospective reach of the \mbox{T-quark} search at the Tevatron Run~II  
superimposed with the bands for which the LTP accounts for all of the 
observed dark matter for three representative values of the Higgs mass, 
140, 250 and 620~GeV.}
\label{fig:DM}
\end{center}
\end{figure}

{\it T-Quark Searches and Dark Matter} --- The relic density of the heavy 
photon LTP in the LHT model is sensitive to its mass, the 
Higgs boson mass, and also to the mass of \mbox{T-quarks} and T-leptons, due to the 
possibility of coannihilation between these states and the LTP.
Assuming that the LTP accounts for all of the dark matter, the precise
measurement of the present dark matter abundance by the WMAP
collaboration ($\Omega_{\rm dm}h^2=0.104^{+0.007}_{-0.010}$~\cite{WMAP3})
imposes a tight constraint on these parameters~\cite{HM,BNPS}. 
In Fig.~\ref{fig:DM}, the constraints obtained in Ref.~\cite{BNPS} 
(for several values of the Higgs mass) are superimposed onto the Tevatron reach in the 
($\tilde{M}$,$M(\tilde{B})$) plane. Tevatron Run~II experiments are  
already in the position to probe some of the parameter space relevant 
for cosmology. If a signal in the $2j+\met$ channel is seen, the 
observed jet $p_T$ distributions and the total cross section should allow
an approximate determination of $\tilde{M}$ and $M(\tilde{B})$. The LHC
will probe the parameter space further, and, together
with the expected direct measurement of the Higgs boson mass,  
will provide a direct collider test of the LTP dark
matter hypothesis. 

%

{\it Discussion ---} The LHT signature discussed here is complementary to
the previously studied signatures involving direct production of the \mbox{T-odd} 
gauge bosons and the triplet Higgs~\cite{HM}, as well as the T-even and
\mbox{T-odd} partners of the top quark~\cite{HM,top}. The process~\leqn{process}
provides a simple signature with relatively low backgrounds and a 
QCD-strength cross section, making it a very promising channel experimentally. 
On the other hand, note that while naturalness puts rather strong constraints
on the masses of the new states in the gauge and top sectors, the \mbox{T-odd}
quarks could be as heavy as 5 TeV without fine tuning. It is therefore 
important 
to pursue searches in all possible channels to maximize the discovery 
potential. 

It is clear that a $2j+{\rm MET}$ signal will not be a conclusive
signature of the LHT model.  
As noted above, the jet
$p_T$, $\eta$ and $\met$ distributions are identical in the LHT model
and the MSSM with matching spectra. While the production cross section for the 
\mbox{T-quarks} and squarks of the same mass is different, the ambiguity in the 
overall mass scale due to the presence of missing energy does not allow one
to easily discriminate between the two cases based on the overall 
rate~\cite{MP}. Unambiguous discrimination between the LHT and SUSY models
will require a measurement of spin correlations in cascade 
decays~\cite{cascade}, if 
such decays are available, or will have to wait until the experiments at the
International Linear Collider (ILC).

{\it Conclusions ---} Events with acoplanar jets and large missing transverse 
energy provide a promising experimental signature for \mbox{T-quarks} of the LHT
model at hadron colliders. Our study indicates that the 
Tevatron Run~II has an interesting reach in this channel, even using only
the first 310 pb$^{-1}$ of the collected data. A dedicated study by
the CDF and D\O\ collaborations in the context of the LHT model will
be of great interest.

{\it Acknowledgments ---} MP is supported by NSF grant
PHY-0355005. Fermilab is operated by the Universities Research
Association Inc., under contract DE-AC02-76CH03000 with the DOE.

\end{document}